\documentclass[sigconf]{acmart}
\usepackage{graphicx} 
\usepackage{xcolor}
\usepackage{booktabs}
\usepackage{acronym}
\usepackage{amsmath} 
\usepackage{amsthm}
\usepackage{commands}
\usepackage{url}
\usepackage[inline]{enumitem}
\usepackage[switch]{lineno}
\usepackage{uppaal}
\usepackage{balance}
\usepackage{subcaption}
\usepackage{multirow}
\usepackage{makecell}
\usepackage{hyperref}
\usepackage{setspace}
\usepackage{tabularx}
\usepackage{comment}

\usepackage{epigraph}
\usepackage{soul}
\usepackage{tikz}
\usetikzlibrary{calc}

\newif\ifreview
\reviewfalse

\theoremstyle{plain}

\theoremstyle{definition}

\acrodef{as}[AS]{Autonomous System}
\acrodefplural{as}[ASs]{Autonomous Systems}
\acrodef{aa}[AA]{Autonomous Agent}
\acrodefplural{aa}[AAs]{Autonomous Agents}
\acrodef{ta}[TA]{Timed Automaton}
\acrodefplural{ta}[TA]{Timed Automata}
\acrodef{pta}[PTA]{Priced Timed Automaton}
\acrodefplural{pta}[PTA]{Priced Timed Automata}
\acrodef{ptg}[PTG]{Priced Timed Game}
\acrodefplural{ptg}[PTGs]{Priced Timed Games}
\acrodef{spta}[SPTA]{Stochastic Priced Timed Automaton}
\acrodefplural{spta}[SPTA]{Stochastic Priced Timed Automata}
\acrodef{ptmdp}[PTMDP]{Priced Timed Markov Decision Process}
\acrodefplural{ptmdp}[PTMDPs]{Priced Timed Markov Decision Processes}
\acrodef{cps}[CPS]{Cyber-Physical System}
\acrodefplural{cps}[CPSs]{Cyber-Physical Systems}
\acrodef{smc}[SMC]{Statistical Model Checking}
\acrodef{idea}[IDEA]{Identity-Aware Architecture for
Autonomous systems}
\acrodef{abm}[ABM]{agent-based model}

\usepackage[skins,breakable]{tcolorbox}

\usepackage[textwidth=1.9cm,textsize=footnotesize]{todonotes}

\begin{document}

\title{\approachName: Safe and Socially-aware Autonomous Systems}

\author{Livia Lestingi}
\affiliation{%
  \institution{Politecnico di Milano}
  \city{Milan}
  \country{Italy}
}
\author{Amel Bennaceur}
\affiliation{%
  \institution{The Open University}
  \city{Milton Keynes}
  \country{UK}
}
\author{Marcello Bersani}
\affiliation{%
  \institution{Politecnico di Milano}
  \city{Milan}
  \country{Italy}
}
\author{Carlos Gavidia-Calderon}
\affiliation{%
  \institution{The Open University}
  \city{Milton Keynes}
  \country{UK}
}
\author{Anastasia Kordoni}
\affiliation{%
  \institution{Lancaster University}
  \city{Lancaster}
  \country{UK}
}
\author{Mark Levine}
\affiliation{%
  \institution{Lancaster University}
  \city{Lancaster}
  \country{UK}
}
\author{Bashar Nuseibeh}
\affiliation{%
  \institution{The Open University}
  \city{Milton Keynes}
  \country{UK}
}
\author{Matteo Rossi}
\affiliation{%
  \institution{Politecnico di Milano}
  \city{Milan}
  \country{Italy}
}

\renewcommand{\shortauthors}{Lestingi et al.}

\begin{abstract}
Autonomous agents operating in socio-critical settings must coordinate with humans under uncertainty while respecting explicit safety constraints.
Existing approaches either account for social dynamics without formal guarantees or provide formal assurance while abstracting away human behaviour.
We introduce \approachName{}, a formally grounded approach for synthesising socially-aware cooperation strategies with safety guarantees.
The cooperation between humans and the autonomous agent is modelled as a Priced Timed Markov Decision Process, and decision-making is formulated as a cost-bounded reachability problem.
We illustrate the approach using an emergency evacuation scenario.
Initial experimental evidence demonstrates the effectiveness of the approach and highlights the trade-offs between optimisation and safety guarantees.
\approachName{} provides a principled foundation for formally assured, socially-aware decision-making in socio-critical systems.
\end{abstract}


\maketitle

\acresetall
\section{Introduction}
\label{sec:intro}

Autonomous agents often require interaction and cooperation with humans to achieve their goals. Such systems are considered safety-critical since they raise the need for guarantees that, even following autonomous agents' decisions,  the system will satisfy safety properties.
However, human behaviour is uncertain and complex. 
As a result, formally reasoning about human behaviour and specifying, designing, and deploying autonomous agents able to cooperate with humans while guaranteeing safety is challenging~\cite{cacm23}.

Many existing approaches model human behaviour probabilistically or as a reaction to system stimuli,
enabling autonomous agents to reason under uncertainty~\cite{EskinsS11,camara2015reasoning,BobuSFSD20}.
However, these approaches typically provide limited guarantees about whether safety constraints
will hold when decisions are enacted in the presence of humans.
For autonomous agents to be able to interact and cooperate with humans, they need to reason about the social structures that drive human behaviour~\cite{cacm23}.
%
\ac{idea}~\cite{tosem24} proposes a software architecture
that enables humans and autonomous agents to cooperate by leveraging the notion of \emph{social identity}~\cite{tajfel2010social}
(\ie that humans tend to behave according to the values and expectations imposed by their social groups when these 
become salient). 
\ac{idea} demonstrates how social identity can be leveraged to support cooperation between humans and autonomous agents.
However, in \ac{idea}, the strategies are derived through equilibrium reasoning and do not provide formal
guarantees that safety constraints---such as time bounds or prioritisation requirements---will always be satisfied.

This paper introduces \approachName{} (Formally-verified Identity-aware Autonomous Agents), an approach
for supporting cooperation between humans and autonomous agents while providing explicit safety guarantees.
The core idea explored in this paper is to synthesise decision-making strategies for autonomous agents that
are both socially-aware and formally verified.
We model human--autonomous agent interaction as a \ac{ptmdp}~\cite{stratego15}, where uncertainty captures socially-driven human responses, timing constraints capture urgency, and prices represent performance objectives.
Safety requirements are formulated as reachability properties, enabling the automated synthesis
of strategies that optimise performance while guaranteeing safety by construction.
By formulating socially-aware cooperation as a cost-bounded reachability problem over \acp{ptmdp}, our approach
differs from prior game-theoretic and probabilistic models by explicitly constraining strategy synthesis with
formal safety requirements.

This paper focuses on the foundational modelling and strategy synthesis principles underlying \approachName.
We present an emergency evacuation case study and initial experimental evidence that illustrates the effectiveness of the approach and the trade-offs between safety guarantees and performance metrics.
A comprehensive evaluation of scalability, deployment constraints, and broader empirical validation is left to ongoing and future work.
Our contributions are as follows:
\begin{itemize}[leftmargin=*]
    \item a \emph{formalisation} of socially-aware human--autonomous agent cooperation through \acp{ptmdp},
    \item a \emph{strategy synthesis} approach that integrates social identity uncertainty with explicit safety guarantees as a cost-bounded reachability problem over \acp{ptmdp}, and 
    \item initial \emph{experimental evidence} demonstrating its effectiveness.
\end{itemize}

The rest of the paper is structured as follows.
Section~\ref{sec:domain} presents the case study.
Section~\ref{sec:bg} outlines preliminary concepts. 
Section~\ref{sec:approach} details the \approachName{} framework.
Section~\ref{sec:eval} reports on the empirical validation.
Section~\ref{sec:rw} reviews related work.
Finally, Section~\ref{sec:conclusion} concludes the paper and outlines the research agenda.

\section{Running Example: Emergency Evacuation}
\label{sec:domain}

\approachName{} targets \emph{socio-critical systems}, in which autonomous agents must coordinate with humans under uncertainty while respecting explicit safety constraints.
\approachName's target applications are characterised by \emph{decision points} at which multiple cooperative actions are available and the choice among them may affect human safety or the use of scarce resources~\cite{baresi2024conceptual}.

Mass emergencies---such as fires, earthquakes, or terrorist attacks---require rapid and coordinated response to minimise harm.
Alongside professional emergency services (\emph{first responders}), ordinary people (\emph{zero responders}) often play a crucial role in early response, particularly when professional resources are scarce~\cite{drury2019facilitating}.
Social identity dynamics may promote pro-social behaviour among survivors, but also introduce uncertainty, as not all individuals are equally willing or able to cooperate.
We consider evacuation scenarios in which an autonomous agent acts as an active participant, supporting coordination between humans and professional responders.
The autonomous agent must make localised decisions at runtime, such as assigning a rescue task to a survivor or contacting a first responder, while ensuring that safety requirements are not violated.

Consider a search-and-rescue robot operating in a disaster zone. When the robot encounters a fallen person who cannot move independently, it must decide how to coordinate assistance.
Possible actions include requesting help from a nearby survivor or contacting a first responder.
The decision implies a trade-off between response time, availability of resources, and uncertainty about human compliance.
Safety requirements may impose constraints such as maximum 
response time or prioritisation of vulnerable individuals.
This example captures the challenges addressed by \approachName{}: cooperation between an autonomous agent and humans, uncertainty arising from socially-driven behaviour, the presence of scarce resources, and urgency imposed by safety-critical conditions.

\section{Preliminaries}
\label{sec:bg}

This section introduces the minimal formal background required to understand how \approachName{} models and synthesises socially-aware, safety-constrained decisions.
We omit full formal definitions for brevity and focus on the core modelling and synthesis concepts.

Socio-critical decision-making problems require reasoning about four intertwined aspects:
\begin{enumerate*}[label=(\roman*)]
    \item \emph{control}, as the autonomous agent selects among alternative actions;
    \item \emph{uncertainty}, as human responses cannot be predicted deterministically;
    \item \emph{time}, as actions incur delays and deadlines must be respected; and
    \item \emph{cost}, as decisions affect performance metrics such as evacuation time or
    the use of scarce resources.
\end{enumerate*}
To capture these aspects jointly, \approachName{} models decision-making problems using \acp{ptmdp} \cite{david2014time}.

\begin{figure}[t]
    \centering
    \includegraphics[width=.7\columnwidth]{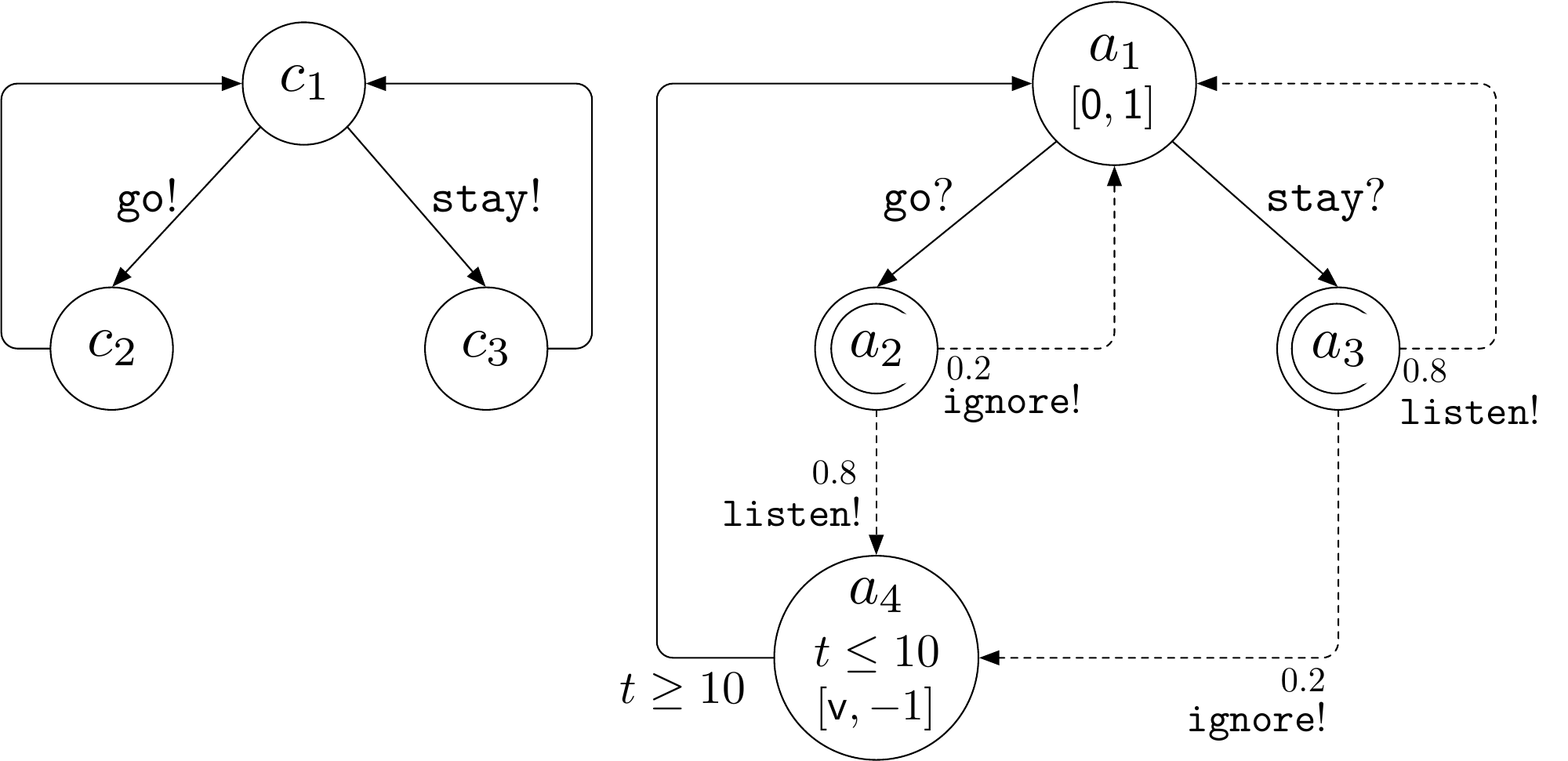}
    \caption{\ac{ptmdp} example. Dashed lines represent uncontrollable edges.
    }
    \label{fig:ptmdpex}
\end{figure}

A \ac{ptmdp} is a stochastic transition system in which an autonomous agent (the \emph{controller}) selects controllable actions, the environment responds probabilistically, and both time and cost accumulate along executions.
Therefore, a \ac{ptmdp} combines:
\begin{enumerate*}[label=(\roman*)]
    \item controllable actions chosen by the autonomous agent;
    \item probabilistic environment responses, capturing uncertain human behaviour;
    \item time constraints governing action durations; and
    \item accumulated costs representing performance objectives.
\end{enumerate*}
Figure~\ref{fig:ptmdpex} shows an illustrative example of a \ac{ptmdp} featuring a controller (the automaton on the left) and the agent reacting to the controller's decisions (the automaton on the right). The controller must decide whether the agent moves or not. In the first case, the agent covers a distance but it spends energy. In the second case, the agent does not move but recovers energy.
Due to uncertainty, there is a 20\% probability that the agent will ignore the controller' decision.

A \ac{ptmdp} consists of a finite set of locations ${\ell \in L}$ representing system states connected by edges ${e\in E}$. Edges are partitioned into controllable (modeling the autonomous agent selecting an action) and uncontrollable (modeling the environment responding probabilistically).
Controllable edges are labeled with actions ${a \in A_c}$, determining the next location.
When an uncontrollable edge fires, the environment selects an action ${a \in A_u}$ according to a probability distribution, modelling uncertain human behaviour.
Each location is annotated with time constraints (\eg ${t<10}$) and a cost vector $\mathbf{c}$ (\eg $[\mathsf{v}, -1]$, where $\mathsf{v}$ denotes the agent's speed), capturing elapsed time or resource usage.

This structure induces executions in which control alternates between the autonomous agent and the environment, while time and cost accumulate along the path.
Strategies resolve controllable choices so as to optimise a cost objective subject to safety constraints.
A strategy is a stochastic function ${\sigma : L \times \mathbb{R}_{\geq 0}\rightarrow\mathcal{D}(A_c)}$,
where $L$ is the set of locations, $\mathbb{R}_{\geq 0}$ represents elapsed time,
$A_c$ is the set of controllable actions, and $\mathcal{D}(A_c)$ denotes a probability
distribution over $A_c$.
\approachName{} synthesises strategies by formulating decision-making as a \emph{cost-bounded reachability problem} \cite{david2014time}.
We denote as $\mathbb{E}[\text{cost}]$ the expected value of a cost metric, and as $\mathcal{P}(\psi)$ the probability that property $\psi$ holds.
Given a \ac{ptmdp}, the goal is to compute a strategy for the autonomous agent that minimises an expected cost while ensuring that a designated safety condition is satisfied.
Formally, we consider strategies $\sigma$ that resolve controllable choices and optimise:
\[
\min \; \mathbb{E}|_{\sigma}[\text{cost}]
\quad \text{subject to} \quad
\mathcal{P}|_{\sigma}\!\left(\Diamond G \;\wedge\; \text{cost} \leq B \right) \geq p ,
\]
where $G$ is a set of goal states satisfying a safety requirement, $B$ is a cost bound (\eg a maximum response time), and $p$ is a required probability threshold.
Operator $\Diamond$ denotes reachability.
This formulation highlights the trade-off between performance optimisation and safety guarantees addressed by \approachName{}.
In the evacuation scenario, $G$ may represent the fallen person receiving assistance within a given
time bound, while the cost captures the duration of the rescue or the involvement of scarce
first-responder resources.

\begin{figure*}[t]
    \centering
    \includegraphics[width=1.5\columnwidth]{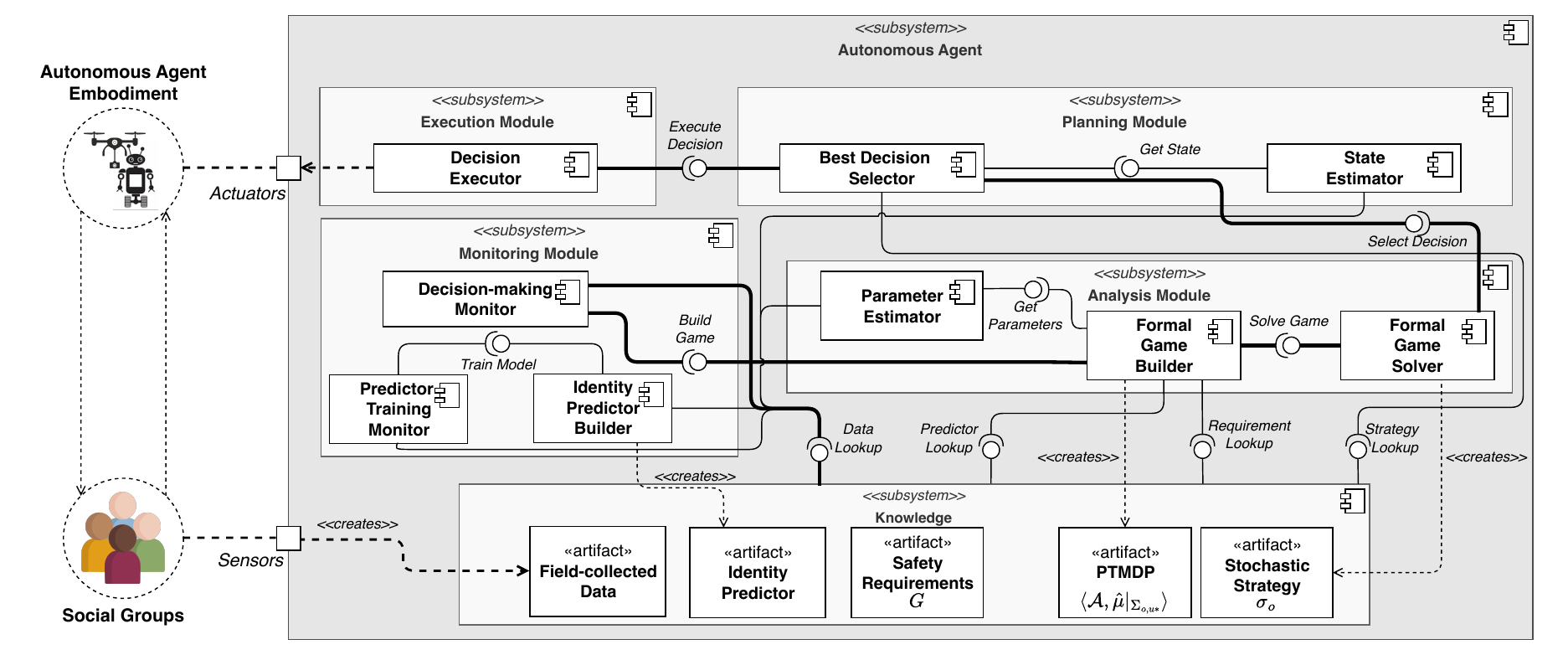}
    \caption{\approachName's architecture (connectors in bold highlight the activation circuit of the main modules).}
    \label{fig:arch}
\end{figure*}

\approachName{} relies on \stratego{} for the synthesis of strategies for \acp{ptmdp} \cite{stratego15}.
\stratego{} provides automated support for computing strategies that optimise quantitative objectives
under probabilistic and timing constraints, as well as \ac{smc} to estimate the likelihood of satisfying safety properties (\ie the value of $\mathcal{P}|_{\sigma}(\Diamond 
G \wedge \text{cost} \leq B)$).

\section{\approachName{}: A Framework for Safe and Socially-aware Autonomous Agents}
\label{sec:approach}

This section presents \approachName{}, focusing on how socially-aware and safety-constrained
decisions are synthesised at runtime. 
Figure~\ref{fig:arch} provides an overview of the \approachName{} framework, illustrating how socially-aware decision-making, formal modelling, and strategy synthesis are integrated.
The figure highlights the main responsibilities of the framework components and their interactions.

\approachName's architecture is structured as a MAPE-K loop with formal strategy synthesis at the core of the analysis and planning phases.
This architectural solution enables \approachName{} to support socially-aware cooperation decisions with explicit safety guarantees without imposing assumptions on perception, actuation, or control mechanisms.
At a high level, \approachName{} operates as a decision-support layer for an autonomous agent.
It is activated when the \comp{Monitoring} module detects a \emph{decision point}, that is, a situation in
which multiple cooperative actions are available and the choice among them may affect safety or performance.
Data collection is assumed to be handled by existing perception components and are treated as external to \approachName{}.

Once invoked, \approachName{} combines three classes of inputs stored in the \comp{Knowledge} subsystem.
First, \emph{field-collected data} describe the current state of the environment, such as distances, availability of resources, or task urgency.
Second, estimates of socially-driven human behaviour are obtained from \emph{identity predictors}, which provide probabilistic information about how humans are likely to
respond to cooperation requests.
Third, \emph{safety requirements} specify the constraints that must be satisfied by any admissible decision.
These inputs are consumed by the \comp{Analysis} module, whose components are responsible for constructing a formal \ac{ptmdp} model representing the current decision context.
The model captures the controllable choices available to the autonomous agent, uncertain human responses, and the relevant time and cost constraints.
Based on this model, the \comp{Formal Game Solver} component computes a strategy that optimises a performance objective while guaranteeing compliance with the specified safety requirements.
The resulting strategy is then used by the \comp{Planning} module to select the best action to be enacted by the \comp{Execution} module.
Low-level control and actuation are outside the scope of \approachName{} and are handled by the autonomous agent itself.
If the environment evolves or new decision points arise, the process can be repeated, leading to the construction of updated models and strategies.

\sloppypar{\textbf{Detecting Decision Points (Monitoring)}.}
\label{sec:mmod}
\approachName{} is invoked when the autonomous agent encounters a decision point, that is, a situation in which multiple cooperative actions are available and the choice may affect safety or performance.
At a decision point, \approachName{} gathers three classes of inputs:
\begin{enumerate*}[label=(\roman*)]
\item contextual information about the current system state (\eg~distances, availability of resources);
\item information about the humans involved, including uncertainty estimates derived from social identity predictors; and
\item safety requirements provided by stakeholders.
\end{enumerate*}
These inputs are treated as parameters for the subsequent formal analysis.

In the evacuation scenario, a decision point occurs when the autonomous agent detects a
fallen person requiring assistance.
The available actions include requesting help from a nearby survivor or contacting a
first responder.
Contextual inputs include distances and response times, while uncertainty arises from the
unknown willingness of a survivor to comply with the request.
Safety requirements may constrain the maximum allowable time before assistance is provided.

\approachName{} relies on an \emph{identity predictor} to estimate the likelihood that a human will comply with a cooperation request.
The identity predictor (stored in the \comp{Knowledge}) takes as input observable identity markers (\eg demographic or
contextual cues such as utterances) and estimates the probability of social identity adoption. 
Therefore, the \comp{Monitoring} module also features a predictor training and update mechanism.
As new interaction data becomes available, the \comp{Predictor Training Monitor} periodically evaluates the accuracy of the current identity predictor.
If performance degradation is detected, a \comp{Identity Predictor Builder} retrains the model using the latest data and updates the predictor stored in the shared knowledge base.

\sloppypar{\textbf{Generating a Verified Optimal Strategy (Analysis).}}
\label{sec:amod}
Given the information collected at a decision point, the \comp{Formal Game Builder} constructs a \ac{ptmdp} that represents the local decision-making context faced by the autonomous agent.
Model construction follows a template-based approach, in which reusable patterns capture recurring aspects of human-autonomous agent cooperation.

\begin{figure}[!t]
\centering
    \begin{subfigure}{.4\columnwidth}
    \centering
        \includegraphics[width=\columnwidth]{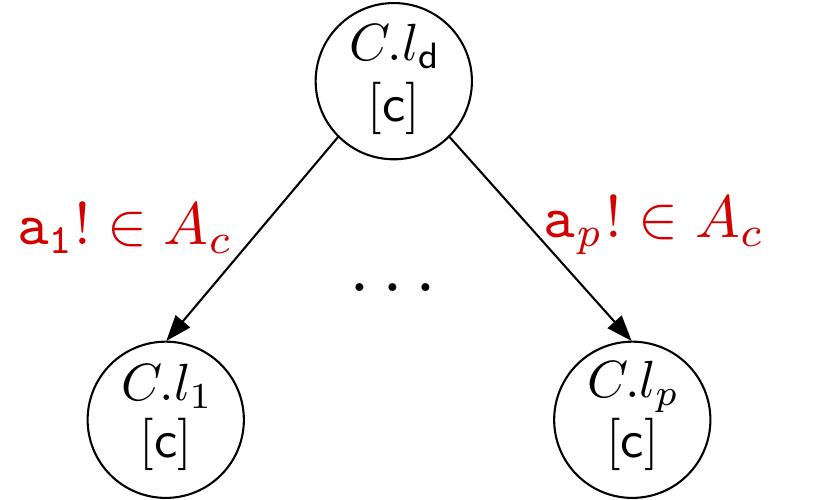}
        \caption{Controller decisions.}
        \label{fig:contract}
    \end{subfigure}
    \begin{subfigure}{.3\columnwidth}
    \centering
        \includegraphics[width=\columnwidth]{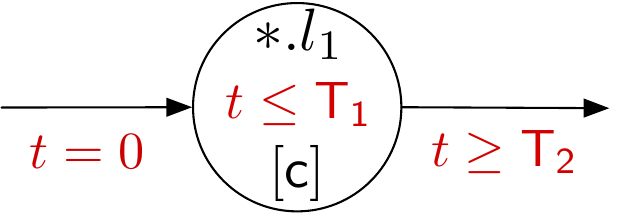}
        \caption{Action implying the elapsing of time.}
        \label{fig:timedact}
    \end{subfigure}
    \begin{subfigure}{.49\columnwidth}
    \centering
        \includegraphics[width=\columnwidth]{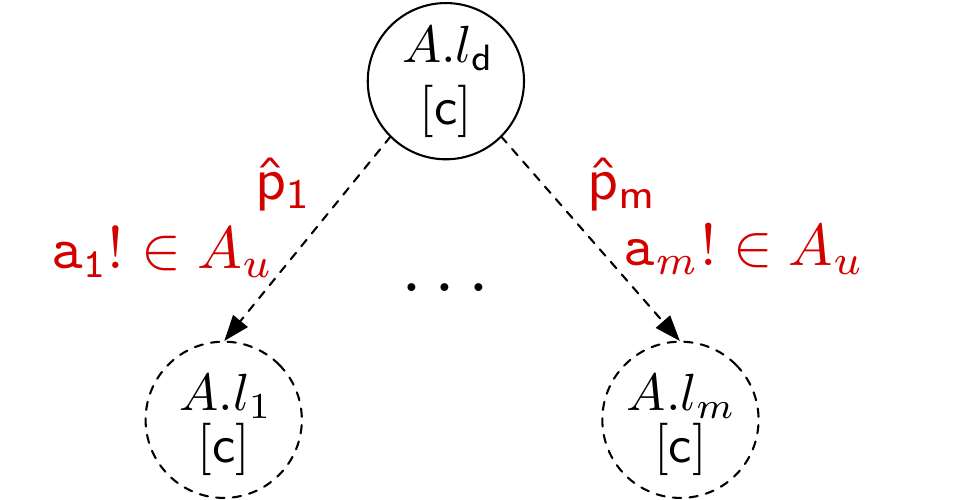}
        \caption{Unobservable adversarial decisions.}
        \label{fig:uncontract}
    \end{subfigure}
    \begin{subfigure}{.49\columnwidth}
    \centering
        \includegraphics[width=\columnwidth]{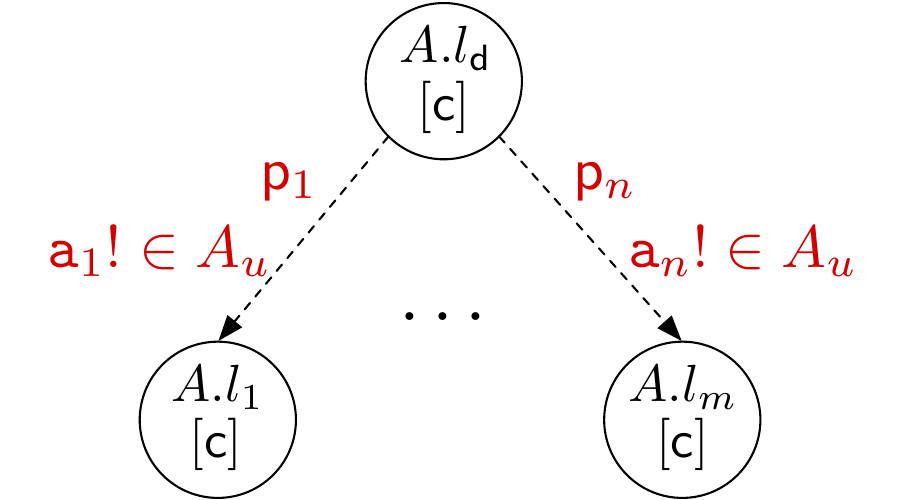}
        \caption{Observable adversarial decisions.}
        \label{fig:obsact}
    \end{subfigure}

    \caption{\ac{ptmdp} patterns composing the controller and adversary models (topical elements of each pattern are in red). 
    Solid lines are controllable edges, while the uncontrollable ones are dashed.}
    \label{fig:patterns}
\end{figure}

Figure~\ref{fig:patterns} depicts the modelling patterns the \comp{Formal Game Builder} uses to construct the controller and environment components of a \ac{ptmdp}.
Each pattern captures a recurring aspect of socio-critical decision-making and can be instantiated and composed to represent a concrete decision context.
Figure~\ref{fig:contract} shows the \emph{controller decision pattern}, in which the autonomous agent selects one action from a finite set of controllable alternatives (\eg assigning a task to a survivor or to a first responder).
These choices correspond to outgoing controllable edges and represent the points at which the autonomous agent exerts control.
Figure~\ref{fig:timedact} models actions that incur a \emph{time overhead}, such as travel or task execution.
The elapsing of time is constrained by parameters (\eg minimum and maximum duration), allowing the model to capture deadlines and urgency.

Figures~\ref{fig:uncontract} and~\ref{fig:obsact} capture uncertainty in human behaviour.
\fref{uncontract} represents \emph{unobservable adversarial decisions}, where the human selects among alternative actions but the outcome is not directly observable by the autonomous agent (\eg whether they have developed a social identity).
Therefore, the humans's biases towards unobservable actions (\ie~weights $\hat{p}_{1\dots m}$) must be estimated through predictors, such as the \comp{Identity Predictor}.
\fref{obsact} represents \emph{observable adversarial decisions}, where the human's choice is visible to the autonomous agent and influences subsequent choices.
Together, these patterns allow \approachName{} to model socially-driven uncertainty without assuming deterministic or perfectly rational behaviour.
A complete \ac{ptmdp} model is obtained by composing these patterns for the controller and one or more environment entities.
Time constraints and cost rates are instantiated from contextual parameters obtained through perception and domain knowledge (\ie through the \comp{Parameter Estimator} component).
Such parameters include, for example, distances between agents, expected task durations, and resource usage costs.
More generally, the construction process yields a \emph{parametric} \ac{ptmdp}, where a set of parameters $k_1,\dots,k_n$ determines the size and quantitative characteristics of the model.
By varying these parameters, the same modelling workflow can be used to generate different instances of a decision problem.

Safety requirements are encoded as reachability goals $G$ and cost bounds $B$, as introduced in Section~\ref{sec:bg}.
Intuitively, $G$ characterises the set of admissible outcomes (\eg successful task completion or prioritisation of vulnerable individuals), while $B$ constrains the maximum acceptable cost (\eg response time or number of scarce resource usages).
The resulting \ac{ptmdp} thus represents a finite, parameterised abstraction of the decision-making problem at hand.

The \comp{Formal Game Solver} generates a strategy by solving a cost-bounded reachability problem over the constructed \ac{ptmdp}.
Using \stratego{}, a strategy $\sigma$ is computed that minimises the expected cost while ensuring that the probability of satisfying the safety condition meets a threshold.
Because safety constraints are embedded directly in the synthesis problem, any decision selected according to $\sigma$ is guaranteed to comply with the specified safety requirements under the assumptions encoded in the model.

\sloppypar{\textbf{Selecting and Executing the Best Decision (Planning and Execution).}}
\label{sec:pmod}
Once a strategy has been synthesised, the \comp{Planning} module uses it to select the action to be enacted by the autonomous agent.
The \comp{State Estimator} identifies the location of the \ac{ptmdp} that most closely represents the state of the physical system and the elapsed time.
Given this pair in $L\times\mathbb{R}_{\geq0}$, $\sigma$ then prescribes a probability distribution over controllable actions.
In practice, the \comp{Best Decision Selector} picks the action with the highest probability (\ie minimum expected cost) and the \comp{Decision Executor} maps it to an executable command.

In the evacuation scenario, this corresponds to issuing a request to a survivor or contacting a first responder.
If the environment evolves or new information becomes available, the process can be repeated at subsequent decision points, yielding updated strategies.

\section{Evaluation}
\label{sec:eval}

This section reports on the preliminary empirical evaluation of the \approachName{} framework applied to the mass emergency running example. 
The experiments evaluate \approachName's \emph{effectiveness} in terms of how the generated strategies improve the performance while guaranteeing safety properties. 
We use two simulation environments for strategies generated through the \approachName{} framework and compare them to four baseline conditions, (1) no autonomous agent supporting the emergency response (\emph{no-support}), (2) always calling the first responders (\emph{staff-support}), (3) always asking zero responders to help each other (\emph{survivor-support}), and (4) an autonomous agent enacting decisions based on strategies generated by IDEA.
Specifically, we address the following questions:
\begin{enumerate}[label=\textbf{RQ\arabic*.}, leftmargin=1.1cm]
    \item How effective are \approachName-generated strategies when deployed in an independently-developed simulation and how does \approachName{} compare against the baselines?
    \item How much does \approachName{} support different safety properties and generate optimised strategies when deployed in \stratego?
    \item What is the time overhead induced by \approachName?
     \item How generalisable is \approachName{} to socio-critical systems featuring the required factors?
\end{enumerate}

\subsection{RQ1: \approachName{} in Action}

\begin{figure}[!t]
\centering
    \begin{subfigure}{\columnwidth}
        \centering
        \includegraphics[width=\columnwidth]{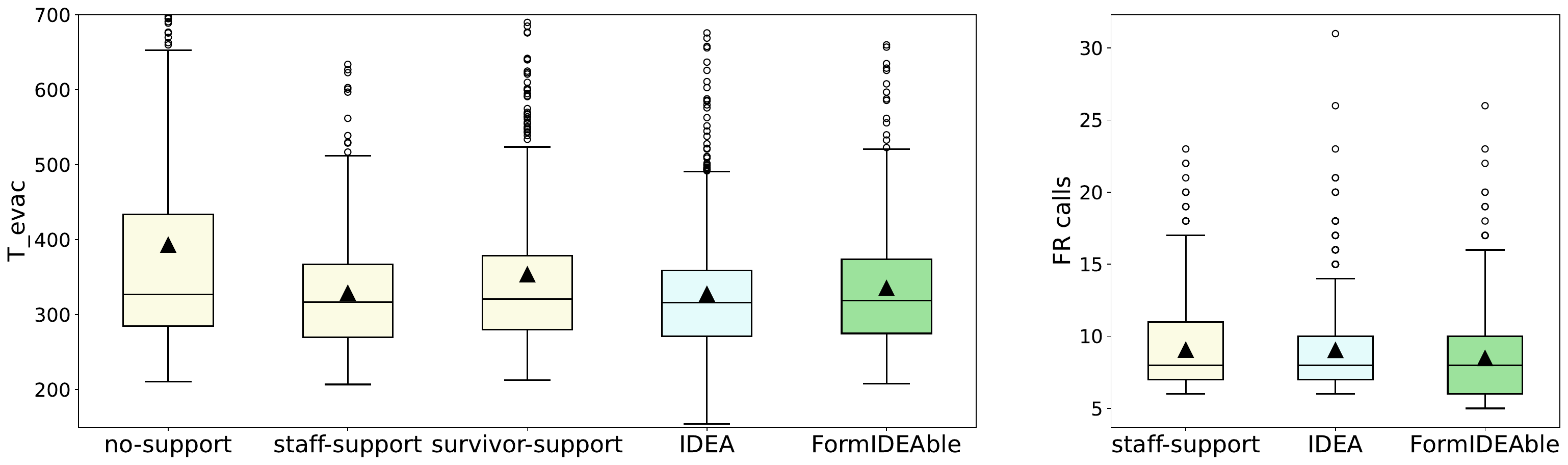}
        \caption{$\mathsf{UB}=10, \mathsf{N_p}\in[150, 700], \mathsf{N_{fr}}=[1, 30]$}
        \label{fig:rq1_1}
    \end{subfigure}
    \begin{subfigure}{\columnwidth}
        \centering
        \includegraphics[width=\columnwidth]{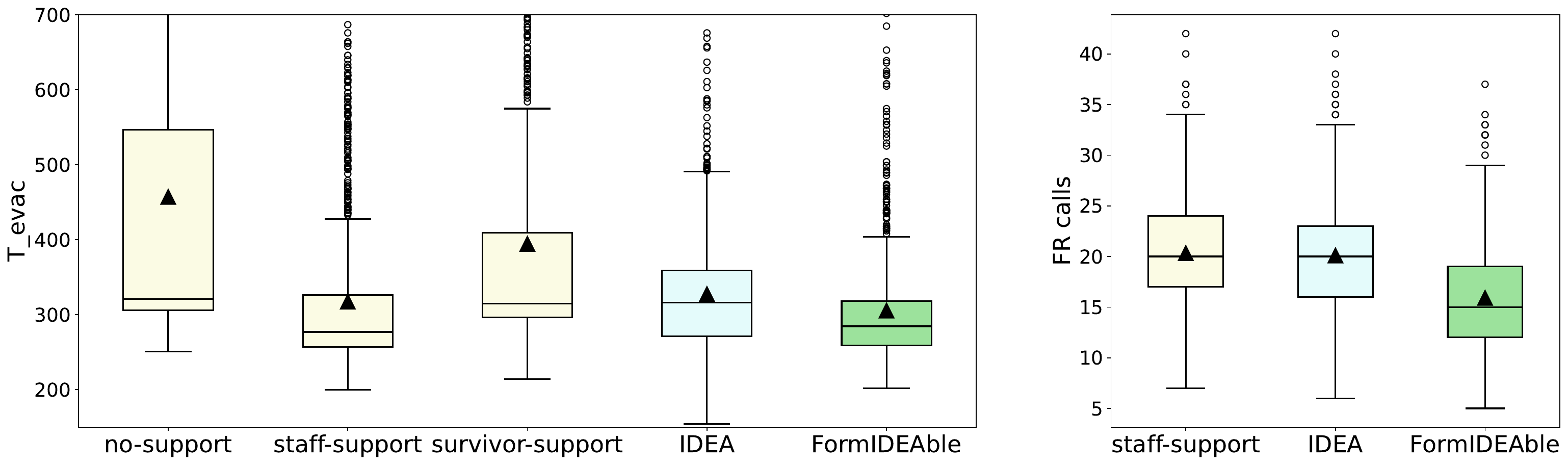}
        \caption{$\mathsf{UB}=10, \mathsf{N_p}=800, \mathsf{N_{fr}}=9$}
        \label{fig:rq1_2}
    \end{subfigure}
\caption{Evacuation time ($\mathsf{T_{evac}}$) and first-responder calls (\textsf{FR\_calls}) distributions obtained with \approachName{} (green), \ac{idea} (blue), and the three baselines (yellow).}
\label{fig:rq1}
\end{figure}

\begin{table}[t]
    \centering
    \caption{Statistical difference (p-value, effect size) between evacuation times ($\mathsf{T_{evac}}$) obtained with \approachName{} compared to \ac{idea} and the baselines.}
    \label{tab:rq1_stat_tevac}
    \resizebox{\columnwidth}{!}{
    \begin{tabular}{c c | c c c c }
    \multirow{2}{*}{$\mathsf{N_{p}}$} & \multirow{2}{*}{$\mathsf{N_{fr}}$} & \multicolumn{4}{c}{\approachName{} vs.} \\
        \cline{3-6}
        & & \textsf{no-support} & \textsf{staff-support} & \textsf{survivor-support} & \textsf{\ac{idea}} \\
        \hline
        $[150, 700]$ & $[1, 30]$ & $\mathbf{<0.05}$ \textbf{(S)} & 1.7E-01 (N) & $\mathbf{<0.05}$ (N) & $\mathbf{<0.05}$ (N) \\
        $800$ & $9$ & $\mathbf{<0.05}$ \textbf{(L)} & 9.0E-01 (N) & $\mathbf{<0.05}$ \textbf{(M)} & $\mathbf{<0.05}$ \textbf{S} \\
    \end{tabular}}
\end{table}

\begin{table}[t]
    \centering
    \caption{Statistical difference (p-value, effect size) between first-responder calls (\textsf{FR\_calls}) obtained with \approachName{} compared to \ac{idea} and the staff-support baseline.}
    \label{tab:rq1_stat_frcalls}
    \resizebox{.6\columnwidth}{!}{
    \begin{tabular}{c c | c c }
    \multirow{2}{*}{$\mathsf{N_{p}}$} & \multirow{2}{*}{$\mathsf{N_{fr}}$} & \multicolumn{2}{c}{\approachName{} vs.} \\
        \cline{3-4}
        & & \textsf{staff-support} & \textsf{\ac{idea}} \\
        \hline
        $[150, 700]$ & $[1, 30]$ & $\mathbf{<0.05}$ (N) & $\mathbf{<0.05}$ (N) \\
        $800$ & $9$ & $\mathbf{<0.05}$ \textbf{(M)} & $\mathbf{<0.05}$ \textbf{(M)} \\
    \end{tabular}}
\end{table}

\begin{table}[t]
    \centering
    \caption{\approachName's accuracy when predicting whether the safety goal will hold compared against in-simulation observations.}
    \label{tab:rq1_precision}
    \resizebox{.55\columnwidth}{!}{
    \begin{tabular}{c c | c | c c}
    $\mathsf{N_{p}}$ & $\mathsf{N_{fr}}$ & $\mathsf{UB}$ & Precision & Recall \\
    \hline
    \multirow{2}{*}{$[150, 700]$} & \multirow{2}{*}{$[1, 30]$} & 5 & \textbf{0.932} & \textbf{0.940} \\
    & & 10 & \textbf{0.990} & \textbf{1.000} \\
    \hline
    \multirow{2}{*}{$800$} & \multirow{2}{*}{$9$} & 5 & \textbf{0.967} & \textbf{0.969} \\
    & & 10 & \textbf{0.997} & \textbf{1.000}
    \end{tabular}}
\end{table}

The goal of this research question is to assess whether strategies synthesised by
\approachName{} are effective when deployed in an independently-developed simulator,
and to compare their performance against \ac{idea}.
To this end, we evaluate \approachName{} in IMPACT+, an agent-based model of emergency
evacuation scenarios~\cite{tosem24}.
IMPACT+ features a rescue robot that, upon detecting a fallen person, must decide whether
to request assistance from a nearby survivor or to contact a first responder.
The simulator is parameterised by the number of evacuees ($\mathsf{N_p}$), the number of
available first responders ($\mathsf{N_{fr}}$), and the maximum time
$\mathsf{T_{fall}}$ a fallen person can wait before resuming evacuation autonomously.

To ensure a fair comparison, both \approachName{} and \ac{idea} rely on the same
\comp{Identity Predictor} to estimate the probability that a survivor has developed a shared
identity.
The predictor is trained on interaction data collected from 100 IMPACT+ simulations and
takes as input identity markers of both the survivor and the victim, yielding an estimate
$\hat{p}_{si}$ of shared identity adoption.

We deploy synthesised strategies using 20 values of $\mathsf{T_{fall}} \in [30,600]$ across two scenarios.
The first samples $\mathsf{N_p} \in [150,700]$ and $\mathsf{N_{fr}} \in [1,30]$ uniformly, while the second fixes $\mathsf{N_p}=800$ and $\mathsf{N_{fr}}=9$, following the experimental setup of \ac{idea}.
Overall, this results in 4000 simulation runs.
For each run, we measure the total evacuation time $\mathsf{T_{evac}}$ and the number of first-responder calls (\textsf{FR\_calls}), comparing \approachName{} against three baselines (no-support, staff-support, survivor-support) and \ac{idea} under identical simulation conditions.

Statistical significance is assessed using the Mann--Whitney U test, with effect sizes reported via Vargha and Delaney’s $\hat{A}_{AB}$ statistic \cite{mannwhitney,vargha}.
We additionally evaluate \approachName{}’s ability to predict whether a safety requirement will hold.
Specifically, we compare the time predicted by the \ac{ptmdp} model against the actual in-simulation time required to provide assistance, reporting precision and recall for different safety bounds $\mathsf{UB}$.

\begin{table*}[t]
    \centering
    \caption{\approachName{} configurations selected to address RQ2. While the approach is agnostic with respect to the application domain, descriptors (\ie \emph{distance} and \emph{duration}) are tailored to the evacuation example.}
    \label{tab:rq2}
    \resizebox{1.9\columnwidth}{!}{
    \renewcommand{\arraystretch}{1.5}
    \begin{tabular}{l l l l l l l}
        Configuration & $N_s$ & $N_v$ & \makecell[l]{Parameters\\ ($k_{1..n}$)} & Parameter Range ($\subset\mathbb{N}$) & Safety Goal ($G$) & Reachability Property ($\psi$) \\
        \hline
        \multirow{2}{*}{\textsf{one-shot}} & \multirow{2}{*}{$1$} & \multirow{2}{*}{$2$} & survivor-to-victim distance & $[1, 20]$ & Task completion & Task completion and child prioritization \\
        \cline{4-7}
        & & & survivor-to-victim distance & $[1, 20]$ & Child prioritization & Task completion and child prioritization \\
        \hline
        \multirow{2}{*}{\textsf{with-memory}} & \multirow{2}{*}{$5$} & \multirow{2}{*}{$5$} & \makecell[l]{survivor-to-victim distance} & $\{1, 11, 21\}$ & \multirow{2}{*}{Task completion and cap on first-responder calls} & \multirow{2}{*}{Task completion and action duration within threshold} \\
        & & & \makecell[l]{maximum rescue action duration \textsf{UB}} & $\{50,40,30\}$ & & \\
    \hline    
    \end{tabular}}
\end{table*}

\paragraph{Results}

Figure~\ref{fig:rq1} reports the distributions of evacuation time and first-responder calls, while Tables~\ref{tab:rq1_stat_tevac} and~\ref{tab:rq1_stat_frcalls} summarise statistical
differences with respect to the baselines.
Across configurations, \approachName{} significantly outperforms the no-support baseline.
When first responders are scarce ($\mathsf{N_p}=800$, $\mathsf{N_{fr}}=9$), \approachName{} also shows statistically significant improvements over \ac{idea} and
survivor-support in evacuation time, while remaining comparable to staff-support.
\approachName{} consistently reduces first-responder involvement compared to both staff-support and \ac{idea}.

Table~\ref{tab:rq1_precision} reports the accuracy of \approachName{} in predicting whether
the safety requirement can be satisfied.
Precision and recall are always above $93\%$, indicating that the formal model provides a
reliable estimate of safety satisfaction despite being deployed in an independent simulator.

\subsection{RQ2: Trade-off between optimality and safety guarantees}

\begin{figure}[!t]
\centering
    \begin{subfigure}{.6\columnwidth}
    \centering
        \includegraphics[width=\columnwidth]{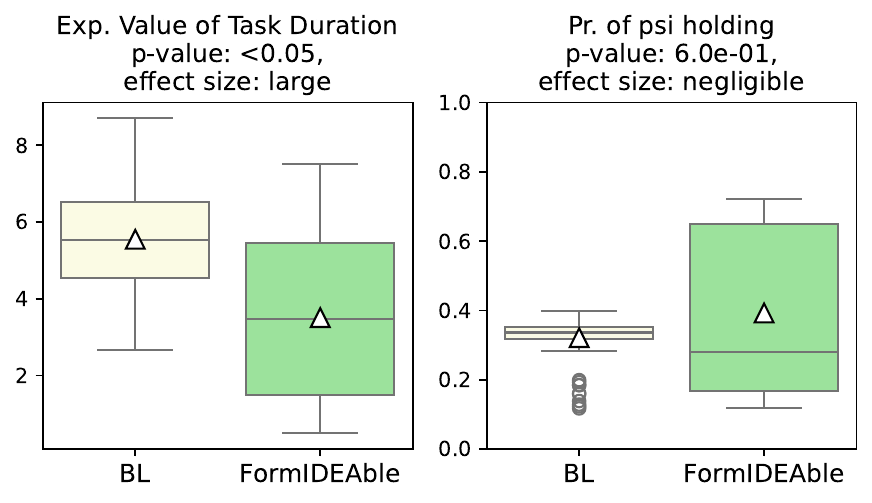}
        \caption{$G:\ \bigvee_{i=1}^{6} A.l_{end,i}$. 
        }
        \label{fig:rq2_2a}
    \end{subfigure}
    \begin{subfigure}{.6\columnwidth}
    \centering
        \includegraphics[width=\columnwidth]{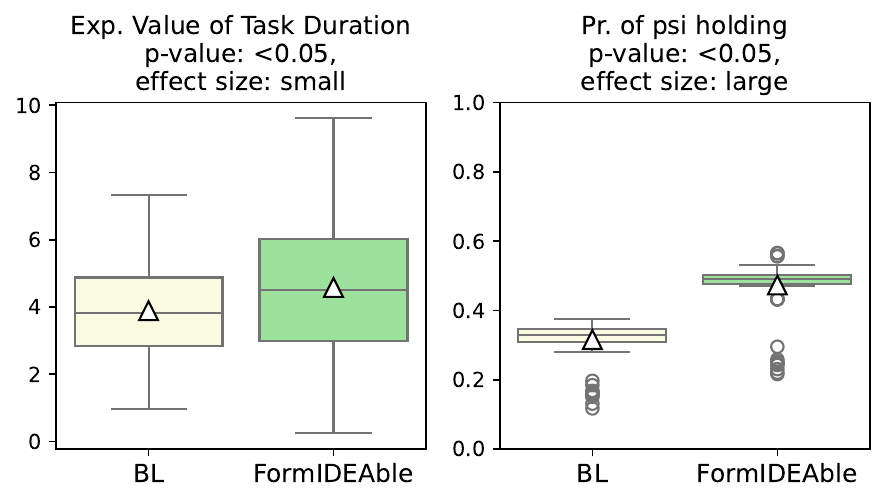}
        \caption{$G:\ \text{v}.\mathit{age}=\mathsf{child}$. 
        }
        \label{fig:rq2_2b}
    \end{subfigure}
\caption{\textsf{\small One-shot} configuration results (${\mathsf{N_s}=1}$, ${\mathsf{N_v}=2}$).}
\label{fig:rq2_1}
\end{figure}

\begin{figure}[!t]
\centering
    \begin{subfigure}{\columnwidth}
        \centering
        \includegraphics[width=.5\columnwidth]{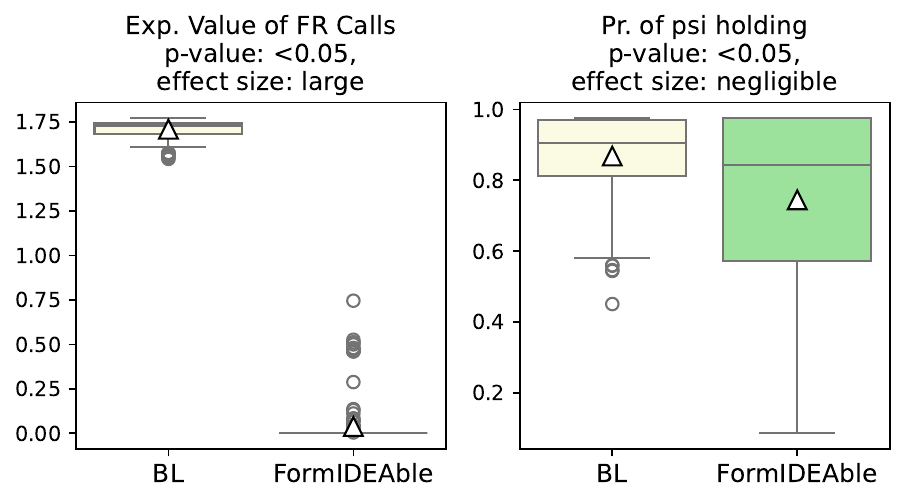}
        \caption{${\mathsf{UB}=50}$}
        \label{fig:rq2_3d}
    \end{subfigure}
    \begin{subfigure}{\columnwidth}
    \centering
        \includegraphics[width=.5\columnwidth]{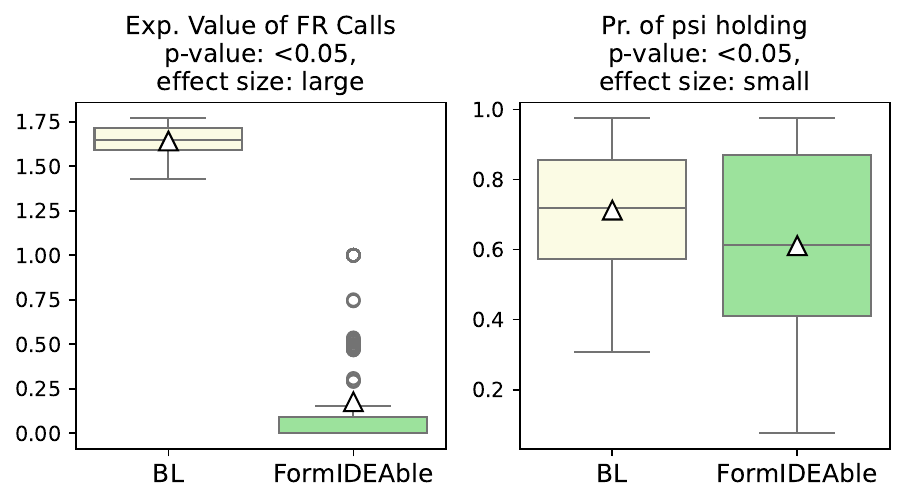}
        \caption{${\mathsf{UB}=40}$}
        \label{fig:rq2_3e}
    \end{subfigure}
    \begin{subfigure}{\columnwidth}
    \centering
        \includegraphics[width=.5\columnwidth]{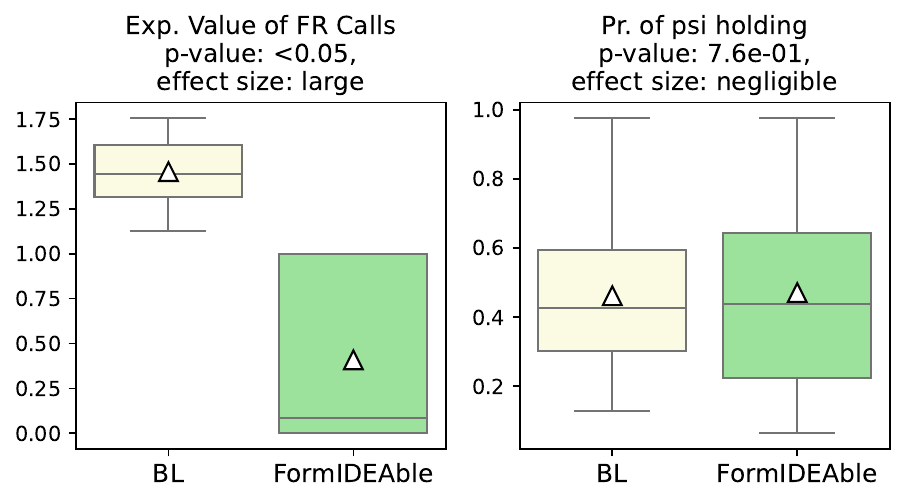}
        \caption{${\mathsf{UB}=30}$}
        \label{fig:rq2_3f}
    \end{subfigure}
\caption{\textsf{With-memory} configuration results (${\mathsf{N_s}=5}$, ${\mathsf{N_v}=5}$).}
\label{fig:rq2_3}
\end{figure}

This research question evaluates \approachName{}’s ability to synthesise strategies under different specifications and decision-making contexts, beyond what can be represented
in IMPACT+.

\paragraph{Experiment Design}

Table~\ref{tab:rq2} summarises the configurations used to evaluate \approachName.
Each configuration is obtained by instantiating the modelling patterns with different numbers of survivors ($N_s$), victims ($N_v$), and contextual parameters.
In the \textsf{one-shot} configuration, a single decision is made to assist one of two victims when only one survivor is available.
Model parameters vary the distance between the survivor and the victim, which affects both task duration and cost.
%
In the \textsf{with-memory} configuration, multiple rescue tasks must be assigned over time with multiple survivors and victims.
In addition to varying distances and time bounds, this configuration introduces a safety requirement that limits the number of first-responder interventions.
The reachability property measures the probability that all rescue tasks are completed within a given time bound while respecting this constraint.
For each configuration and point in the configuration's parameter space, \approachName{} synthesises a strategy $\sigma$ that minimises task duration while guaranteeing a given safety goal $G$.
In the \textsf{one-shot} configuration, we consider goals that require task completion or prioritisation of a child victim.
In the \textsf{with-memory} configuration, the goal additionally constrains the number of first-responder calls.
In all cases, \ac{smc} is used to estimate the probability that a reachability property $\psi$ holds under the synthesised strategy.
All strategies are evaluated in \stratego{} and compared against a non-strategised baseline (BL) in which controllable actions are selected randomly with a uniform distribution.
We assess the impact of strategy synthesis by comparing:
\begin{enumerate*}
    \item expected values of cost metrics and
    \item the probability of satisfying the reachability property $\psi$,
\end{enumerate*}
using the Mann--Whitney U test and Vargha and Delaney’s $\hat{A}_{AB}$ effect size.

\paragraph{Results}
Figures~\ref{fig:rq2_1} and~\ref{fig:rq2_3} summarise the results.
In the \textsf{one-shot} setting ($\mathsf{N_s}=1$, $\mathsf{N_v}=2$), \approachName{} significantly reduces task duration when only termination is required, but does not consistently improve the probability of rescuing the child.
When child prioritisation is explicitly enforced as a safety goal, the synthesised strategy substantially increases the probability of satisfying $\psi$, at the cost of a modest increase in task duration.
In the \textsf{with-memory} configuration, 
\approachName{} consistently reduces the number of first-responder calls compared to the
baseline across all safety bounds.
As the time constraint becomes stricter, the probability of satisfying $\psi$ decreases for both approaches; however, \approachName{} maintains comparable safety satisfaction while significantly reducing reliance on scarce resources (see
Figures~\ref{fig:rq2_3d}--\ref{fig:rq2_3f}). 

\subsection{RQ3: Time Overhead}

This research question aims at assessing the wall-clock time overhead induced by the deployment of \approachName.

\paragraph{Experiment Design}
The analysis is carried out by clocking the time necessary to complete a \approachName{} run with the configurations selected for RQ1 and RQ2.
In this case, the alternative to deploying \approachName{} is having the autonomous agent employ no strategy or choose an action randomly (serving as the baseline), for which the associated time overhead is approximated to $0$.

Experiments are performed on a machine running macOS Sonoma 14.5 with an Apple M3 processor and 24GB of memory. Verification is performed through \uppaal v.5.1.0.

\paragraph{Results}

\begin{figure}[t]
    \centering
    \includegraphics[width=.8\columnwidth]{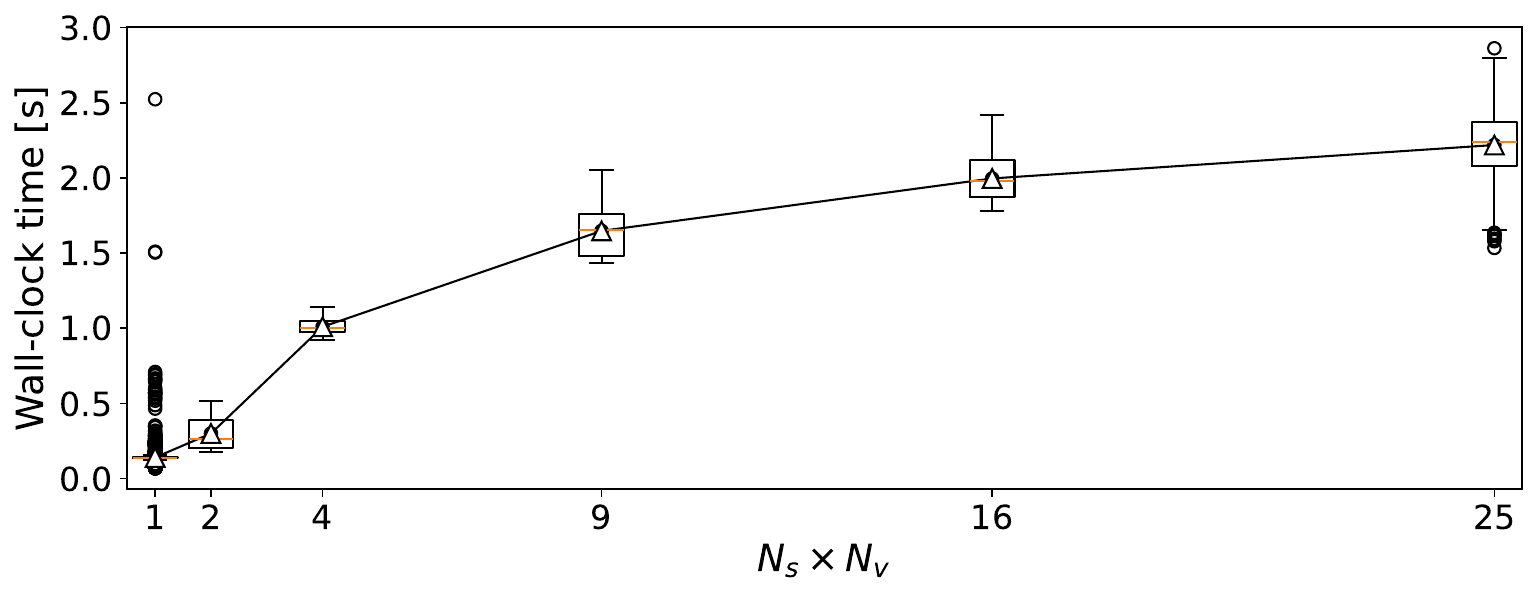}
\caption{Wall-clock time for RQ1 and RQ2 experiments for increasingly complex models ($\mathsf{N_s}\times\mathsf{N_v}$).}
\label{fig:rq3}
\end{figure}

Figure~\ref{fig:rq3} reports the wall-clock time required to synthesise strategies for the configurations used in RQ1 and RQ2, with increasing \approachName{} model size.
Model size is measured by the number of individuals considered by the autonomous agent, expressed as $\mathsf{N_s}\times\mathsf{N_v}$.
Across all configurations, the synthesis overhead introduced by \approachName{} is non-negligible compared to a zero-cost baseline.
However, the overhead remains modest for small and medium-sized configurations, with average runtimes below $0.5$s for up to two individuals and below $0.15$s for single-individual
decisions.
The highest observed overhead (approximately $2.25$s on average) corresponds to the most complex configurations explored in RQ2.
These results indicate that \approachName{} is suitable for scenarios where safety-critical decisions allow for limited deliberation time.
Further scalability optimisations are left to future work.

\subsection{RQ4: Generalisability}
This research question examines the generalisability of \approachName{} by grounding it in established evidence that social identity shapes cooperative behaviour across domains such as security, privacy, and healthcare~\cite{DBLP:conf/icse/RaufLLTNR20,DBLP:conf/icse/CalikliLBRDPSLN16,stuart2022loneliness}. To do so, we move from emergency evacuations as a running example and qualitatively analyse three distinct socio-technical domains to explore how \approachName{} can generalise. 
The three case studies focus on healthcare cybersecurity (the Irish Health Service Executive (HSE) ransomware attack)~\cite{HSE2021ContiReport}, community-based care for vulnerable older adults (SERVICE)~\cite{BennaceurSPBLCC23}, and robotics/software engineering (RSE)~\cite{Gavidia-Calderon23}.
For each of these cases, we present the socio-critical factors such as decision points under uncertainty, scarce resources, and fragmented group structures that motivate identity-aware autonomous support and frame future work needed to fully validate the approach.

The three case studies demonstrate how identity dynamics influence coordination, resilience, and safety-critical decision-making. In the HSE case, a lack of clear ownership and preparedness contrasted with strong prosocial behaviour among local hospital teams, revealing how shared crisis identities can emerge and how identity-aware agents could support decentralised yet coordinated responses. In SERVICE, prior work shows that circles of support rely not only on practical assistance but also on psychologically meaningful identities, suggesting that FormIDEAble agents could reason about identity to guide collaborative and safe interventions. Finally, in RSE, identity-based game-theoretic models explain cooperation and sanctioning behaviours in teams such as open-source communities, enabling agents to design nudges that restore collaboration during incidents or failures. Taken together, these cases support the claim that \approachName{} can generalise across socio-critical systems by leveraging shared identities to manage scarce resources, coordinate human–agent collaboration, and uphold safety-relevant properties.

\paragraph{\textbf{Implications for Software Engineering}}
Human-\ac{as} cooperation is increasingly advocated in software engineering whether through Software Engineering 2.0~\cite{Lo23} or AIware~\cite{HassanLRGC00TOL24}.
In this context, human and \ac{as} agents work together to carry out software engineering tasks such as code reviews. 
Achieving this vision requires going beyond automation alone~\cite{LoLLM24} and understanding the behaviour of software practitioners. 
By coordinating group behaviour, \approachName{} lays the foundation for collaboration between \ac{as} and humans.
In particular, emergencies manifest themselves in software engineering through incidents, malicious or otherwise. \ac{as} have a potentially critical role in coordinating the incident response by enabling groups of developers (first responders) and users (zero responders) to collaborate to manage critical incidents (\eg~security incidents). 
We plan to explore the role of \approachName{} in coordinating incident response of multiple groups of developers and users.

\subsection{Threats to Validity}
\label{sec:disc}

This section discusses current limitations 
and the threats to validity
of the reported preliminary results, highlighting directions for future work.
The current version of \approachName{} requires manual effort in two key steps:
\begin{enumerate*}[label=(\roman*)]
\item instantiating and composing \ac{ptmdp} patterns to model a specific socio-critical
decision-making context, and
\item formalising safety requirements as reachability properties.
\end{enumerate*}
While the underlying modelling and synthesis techniques are amenable to automation, the development of higher-level notations or tooling to support these tasks is left for future work.


The empirical evaluation relies on simulation-based studies and formal analysis using \stratego{}.
While this allows controlled experimentation and reproducibility, it constitutes a threat to external validity.
In particular, the behaviour of humans in real-world  emergencies may deviate from the assumptions in IMPACT+. 
To address this limitation, future work will ground the probabilistic models of human behaviour on empirical data collected from real-world scenarios.
In particular, datasets capturing crowd dynamics and human movement under constraints can be leveraged to calibrate and validate the parameters used in the \ac{ptmdp} models.

Furthermore, the evaluation focuses on a single application domain, namely emergency evacuation scenarios.
Although this domain captures key characteristics of socio-critical systems, the results may not generalise directly to other settings without additional modelling effort.
We plan to assess the generalisability of \approachName{} across multiple socio-critical domains that share the same decision-making structure but differ in contextual features.
Finally, scalability results are limited to the range of configurations explored in RQ3; larger models may require further optimisation or abstraction techniques.


\section{Related Work}
\label{sec:rw}

This section discusses related work along three research directions: human--\acp{as} cooperation, trustworthiness of \acp{as}, and safety in human--\acp{as} cooperation.

\sloppypar{\textbf{Human-Autonomous Systems Cooperation.}}
A long-standing challenge in engineering \acp{as} is coping with uncertainty arising from requirements, environments, other systems, and humans~\cite{bertinoro2023}.
This work focuses on uncertainty stemming from human behaviour. In socio-critical systems, humans interacting with autonomous agents are not merely sources of input but first-class participants whose actions directly influence
system decisions and outcomes~\cite{DBLP:conf/seams/PurandareSICC23}.
Several approaches model human behaviour using behaviourist or probabilistic assumptions, treating actions as responses to stimuli~\cite{heimlich2008understanding,camara2015reasoning,DBLP:conf/seams/LiCGSJ21}.
Such models are widely adopted in \emph{human-in-the-loop} systems, where humans intervene at predefined decision points~\cite{DBLP:conf/icse/Lemos20}.
Other work emphasises social adaptation, allowing \acp{as} to update their behaviour based on user feedback or contextual changes~\cite{AlmalikiFBPA14}, including in emergency-management settings~\cite{DBLP:conf/iscram/JohnsonCSMP21}.
With increasing autonomy, \emph{human-on-the-loop} approaches treat humans as strategic supervisors overseeing autonomous planning and execution~\cite{DBLP:journals/concurrency/FischerGJRWR21,
DBLP:conf/icse/LiAKG20}.
While these approaches support coordination, they do not explicitly capture the social structures that shape cooperative human behaviour.

To address this gap, research on joint action~\cite{sebanz2006joint} and human--machine teaming~\cite{grigore:2013,cleland23} emphasises the mutual understanding between humans and autonomous systems.
The \textit{MAPE-K}$_{\textit{HMT}}$ framework leverages runtime models to support bidirectional adaptation between humans and machines.
IDEA~\cite{tosem24} further incorporates social identity theory to reason about cooperation between heterogeneous human groups.
However, neither \textit{MAPE-K}$_{\textit{HMT}}$ nor IDEA provide formal guarantees that cooperative decisions satisfy safety constraints.
\approachName{} builds on social identity theory and game-theoretic interaction models, while additionally synthesising strategies that are safe by construction.

\sloppypar{\textbf{Trustworthiness of Autonomous Systems in Emergencies.}}
Empirical studies show that people tend to follow rescue robots during emergencies, even after observing incorrect robot behaviour~\cite{wagner2015towards,
sakour2016robot,robinette2016overtrust,jiang2019social}.
At the same time, the presence and behaviour of other humans significantly influence trust and compliance~\cite{nayyar2019effective}.
This raises concerns about overtrust and highlights the need for autonomous systems to behave in a manner that is not only trusted but trustworthy~\cite{lee2004trust,cacm23}.
In this context, ensuring that autonomous decisions are socially grounded and aligned with safety requirements is critical.
\approachName{} contributes to this goal by synthesising cooperation strategies that explicitly account for socially-driven uncertainty while guaranteeing safety properties.

\sloppypar{\textbf{Safety of Human-Autonomous Systems Cooperation.}}
Assurance has been a central concern in the development of autonomous systems~\cite{DBLP:journals/corr/abs-1903-04771}, with formal methods playing a key role in guaranteeing requirement satisfaction.
However, shared control and adaptive behaviour in socio-technical systems challenge traditional verification techniques, particularly in the presence of human variability
and uncertainty~\cite{DBLP:journals/cacm/Kress-GazitEHAA21}.
Beyond functional safety, autonomous systems increasingly need to satisfy Social, Legal, Ethical, Empathetic, and Cultural requirements~\cite{cacm23,DBLP:journals/corr/abs-2307-03697}.
Existing approaches address these challenges through alignment mechanisms such as cooperative inverse reinforcement learning~\cite{Hadfield-Menell16} or by weakening human obligations via abstraction and controller synthesis~\cite{TunBN20}.
\approachName{} aims to balance formality and feasibility~\cite{BersaniCLMRS23} by synthesising strategies that optimise quantitative objectives (\eg evacuation time)
while ensuring compliance with explicit safety properties.\\

In summary, \approachName{} lies at the intersection of these three research directions.
It builds on prior work on human-autonomous agent cooperation by explicitly modelling socially-driven uncertainty, draws on insights from trustworthiness research by grounding autonomous decisions in socially-aware reasoning, and advances the state of the art in safety assurance by synthesising cooperation strategies with formal safety guarantees.
Unlike existing approaches, which typically address these dimensions in isolation, \approachName{} integrates them within a single strategy synthesis framework, enabling autonomous systems to coordinate with humans under uncertainty while complying with safety requirements.

\section{Conclusion}
\label{sec:conclusion}

This paper presents \approachName{}, an approach for synthesising socially-aware cooperation strategies with explicit safety guarantees in socio-critical systems.
The core contribution lies in modelling human-\ac{as} interaction as a \ac{ptmdp} and formulating decision-making as a cost-bounded reachability problem, enabling the automated synthesis of strategies that balance performance objectives with safety constraints.
Through simulation-based studies and formal analysis, we provide initial evidence of the feasibility and potential of this approach.

This work establishes the foundational modelling and synthesis principles underlying \approachName{}.
Several extensions are required to develop this work into a mature solution.
Future work includes:
\begin{enumerate*}
    \item deeper empirical validation across additional socio-critical domains, 
    \item systematic scalability analysis of strategy synthesis, and
    \item improved support for modelling efforts, including higher-level specification of decision contexts and safety requirements.
\end{enumerate*}
In addition, future work will investigate how uncertainty in social identity prediction propagates to safety guarantees and how richer classes of requirements beyond safety properties---such as normative requirements---can be integrated.
In conclusion, this paper provides a principled approach for assured, socially-aware decision-making in autonomous agents and outlines a research agenda toward a comprehensive contribution.

\section*{Data Availability}

A replication package is provided at: \href{https://zenodo.org/records/13754285}{https://zenodo.org/records/13754285}.

\section*{Acknowledgements}

This work was supported by the Engineering and Physical Sciences Research Council under grant numbers EP/V026747/1 and EP/R013144/1, and by Science Foundation Ireland under grant number 13/RC/2094\_P2.

\bibliographystyle{ACM-Reference-Format}
\bibliography{refs}

@techreport{HSE2021ContiReport,
  title        = {Conti Cyber Attack on the HSE: Independent Post Incident Review},
  author       = {{PricewaterhouseCoopers (PwC)}},
  institution  = {Health Service Executive (HSE), Ireland},
  type         = {Independent Post Incident Review},
  number       = {HSE Publications},
  year         = {2021},
  month        = dec # "~3",
  url          = {https://www.hse.ie/eng/services/publications/conti-cyber-attack-on-the-hse-full-report.pdf},
  note         = {Commissioned by the HSE Board in conjunction with the CEO and Executive Management Team},
}

@inproceedings{Gavidia-Calderon23,
  author       = {Carlos Gavidia{-}Calderon and
                  Amel Bennaceur and
                  Tamara Lopez and
                  Anastasia Kordoni and
                  Mark Levine and
                  Bashar Nuseibeh},
  title        = {Meet your Maker: {A} Social Identity Analysis of Robotics Software
                  Engineering},
  booktitle    = {{TAS}},
  pages        = {44:1--44:5},
  publisher    = {{ACM}},
  year         = {2023}
}

@inproceedings{BennaceurSPBLCC23,
  author       = {Amel Bennaceur and
                  Avelie Stuart and
                  Blaine A. Price and
                  Arosha K. Bandara and
                  Mark Levine and
                  Linda Clare and
                  Jessica Cohen and
                  Ciaran McCormick and
                  Vikram Mehta and
                  Mohamed Bennasar and
                  Daniel Gooch and
                  Carlos Gavidia{-}Calderon and
                  Anastasia Kordoni and
                  Bashar Nuseibeh},
  title        = {Socio-Technical Resilience for Community Healthcare},
  booktitle    = {{TAS}},
  pages        = {26:1--26:6},
  publisher    = {{ACM}},
  year         = {2023}
}

@article{mannwhitney,
  title={On a test of whether one of two random variables is stochastically larger than the other},
  author={Mann, Henry B and Whitney, Donald R},
  journal={The annals of mathematical statistics},
  pages={50--60},
  year={1947},
  publisher={JSTOR}
}

@article{vargha,
  title={A critique and improvement of the CL common language effect size statistics of McGraw and Wong},
  author={Vargha, Andr{\'a}s and Delaney, Harold D},
  journal={Journal of Educational and Behavioral Statistics},
  volume={25},
  number={2},
  pages={101--132},
  year={2000},
  publisher={Sage Publications Sage CA: Los Angeles, CA}
}

@book{tajfel2010social,
  title={Social identity and intergroup relations},
  author={Tajfel, Henri},
  volume={7},
  year={2010},
  publisher={Cambridge University Press}
}

@inproceedings{stratego15,
  author       = {Alexandre David and
                  Peter Gj{\o}l Jensen and
                  Kim Guldstrand Larsen and
                  Marius Mikucionis and
                  Jakob Haahr Taankvist},
  title        = {Uppaal {Stratego}},
  booktitle    = {Intl. Conf. on Tools and Algorithms for the Construction and Analysis of Systems},
  series       = {Lecture Notes in Computer Science},
  volume       = {9035},
  pages        = {206--211},
  publisher    = {Springer},
  year         = {2015},
  url          = {https://doi.org/10.1007/978-3-662-46681-0\_16},
  doi          = {10.1007/978-3-662-46681-0\_16},
  bibsource    = {dblp computer science bibliography, https://dblp.org}
}

@article{Hadfield-Menell16,
  author       = {Dylan Hadfield{-}Menell and
                  Anca D. Dragan and
                  Pieter Abbeel and
                  Stuart Russell},
  title        = {Cooperative Inverse Reinforcement Learning},
  journal      = {CoRR},
  volume       = {abs/1606.03137},
  year         = {2016},
  url          = {http://arxiv.org/abs/1606.03137},
  eprinttype    = {arXiv},
  eprint       = {1606.03137},
  bibsource    = {dblp computer science bibliography, https://dblp.org}
}

@inproceedings{BobuSFSD20,
  author       = {Andreea Bobu and
                  Dexter R. R. Scobee and
                  Jaime F. Fisac and
                  S. Shankar Sastry and
                  Anca D. Dragan},
  _editor       = {Tony Belpaeme and
                  James E. Young and
                  Hatice Gunes and
                  Laurel D. Riek},
  title        = {{LESS} is More: Rethinking Probabilistic Models of Human Behavior},
  booktitle    = {Intl. Conf. on Human-Robot Interaction},
  pages        = {429--437},
  publisher    = {{ACM}},
  year         = {2020},
  url          = {https://doi.org/10.1145/3319502.3374811},
  doi          = {10.1145/3319502.3374811},
  timestamp    = {Mon, 18 Jul 2022 16:47:42 +0200},
  biburl       = {https://dblp.org/rec/conf/hri/BobuSFSD20.bib},
  bibsource    = {dblp computer science bibliography, https://dblp.org}
}

@inproceedings{EskinsS11,
  author       = {Douglas Eskins and
                  William H. Sanders},
  title        = {The Multiple-Asymmetric-Utility System Model: {A} Framework for Modeling
                  Cyber-Human Systems},
  booktitle    = {Intl. Conf. on Quantitative Evaluation of Systems},
  pages        = {233--242},
  publisher    = {{IEEE} Computer Society},
  year         = {2011},
  url          = {https://doi.org/10.1109/QEST.2011.38},
  doi          = {10.1109/QEST.2011.38},
  bibsource    = {dblp computer science bibliography, https://dblp.org}
}

@article{tosem24,
  title={The {IDEA} of Us: An Identity-Aware Architecture for Autonomous Systems},
  author={Gavidia-Calderon, Carlos and Kordoni, Anastasia and Bennaceur, Amel and Levine, Mark and Nuseibeh, Bashar},
  journal={ACM Trans. on Soft. Engineering and Methodology},
  year={2024},
  publisher={ACM New York, NY}
}

@inproceedings{david2014time,
  title={On time with minimal expected cost!},
  author={David, Alexandre and Jensen, Peter G and Larsen, Kim Guldstrand and Legay, Axel and Lime, Didier and S{\o}rensen, Mathias Grund and Taankvist, Jakob H},
  booktitle={Automated Technology for Verification and Analysis},
  pages={129--145},
  year={2014},
  organization={Springer}
}

@inproceedings{BersaniCLMRS23,
  author       = {Marcello M. Bersani and
                  Matteo Camilli and
                  Livia Lestingi and
                  Raffaela Mirandola and
                  Matteo G. Rossi and
                  Patrizia Scandurra},
  title        = {Towards Better Trust in Human-Machine Teaming through Explainable
                  Dependability},
  booktitle    = {20th International Conference on Software Architecture, {ICSA} 2023
                  - Companion, L'Aquila, Italy, March 13-17, 2023},
  pages        = {86--90},
  publisher    = {{IEEE}},
  year         = {2023},
  url          = {https://doi.org/10.1109/ICSA-C57050.2023.00029},
  doi          = {10.1109/ICSA-C57050.2023.00029},
  bibsource    = {dblp computer science bibliography, https://dblp.org}
}

@article{drury2019facilitating,
  title={Facilitating collective psychosocial resilience in the public in emergencies: Twelve recommendations based on the social identity approach},
  author={Drury, John and Carter, Holly and Cocking, Chris and Ntontis, Evangelos and Tekin Guven, Selin and Aml{\^o}t, Richard},
  journal={Frontiers in public health},
  volume={7},
  pages={141},
  year={2019},
  publisher={Frontiers Media SA}
}

@article{stuart2022loneliness,
  title={Loneliness in older people and COVID-19: applying the social identity approach to digital intervention design},
  author={Stuart, Avelie and Katz, Dmitri and Stevenson, Clifford and Gooch, Daniel and Harkin, Lydia and Bennasar, Mohamed and Sanderson, Lisa and Liddle, Jacki and Bennaceur, Amel and Levine, Mark and others},
  journal={Computers in Human Behavior Reports},
  pages={100179},
  year={2022},
  publisher={Elsevier}
}

@article{jiang2019social,
  title={Social network, activity space, sentiment, and evacuation: what can social media tell us?},
  author={Jiang, Yuqin and Li, Zhenlong and Cutter, Susan L},
  journal={Annals of the American Association of Geographers},
  volume={109},
  number={6},
  pages={1795--1810},
  year={2019},
  publisher={Taylor \& Francis}
}

@inproceedings{robinette2016overtrust,
  title={Overtrust of robots in emergency evacuation scenarios},
  author={Robinette, Paul and Li, Wenchen and Allen, Robert and Howard, Ayanna M and Wagner, Alan R},
  booktitle={2016 11th ACM/IEEE international conference on human-robot interaction (HRI)},
  pages={101--108},
  year={2016},
  organization={IEEE}
}

@inproceedings{TunBN20,
  author       = {Thein Than Tun and
                  Amel Bennaceur and
                  Bashar Nuseibeh},
  _editor       = {Travis D. Breaux and
                  Andrea Zisman and
                  Samuel Fricker and
                  Martin Glinz},
  title        = {{OASIS:} Weakening User Obligations for Security-critical Systems},
  booktitle    = {28th {IEEE} International Requirements Engineering Conference, {RE}
                  2020, Zurich, Switzerland, August 31 - September 4, 2020},
  pages        = {113--124},
  publisher    = {{IEEE}},
  year         = {2020},
  url          = {https://doi.org/10.1109/RE48521.2020.00023},
  doi          = {10.1109/RE48521.2020.00023},
  timestamp    = {Sun, 25 Jul 2021 11:49:37 +0200},
  biburl       = {https://dblp.org/rec/conf/re/TunBN20.bib},
  bibsource    = {dblp computer science bibliography, https://dblp.org}
}

@article{lee2004trust,
  title={Trust in automation: Designing for appropriate reliance},
  author={Lee, John D and See, Katrina A},
  journal={Human factors},
  volume={46},
  number={1},
  pages={50--80},
  year={2004},
  publisher={SAGE Publications Sage UK: London, England}
}

@inproceedings{nayyar2019effective,
  title={Effective robot evacuation strategies in emergencies},
  author={Nayyar, Mollik and Wagner, Alan R},
  booktitle={2019 28th IEEE International Conference on Robot and Human Interactive Communication (RO-MAN)},
  pages={1--6},
  year={2019},
  organization={IEEE}
}

@article{wagner2015towards,
  title={Towards robots that trust: Human subject validation of the situational conditions for trust},
  author={Wagner, Alan R and Robinette, Paul},
  journal={Interaction studies},
  volume={16},
  number={1},
  pages={89--117},
  year={2015},
  publisher={John Benjamins}
}

@inproceedings{sakour2016robot,
  title={Robot assisted evacuation simulation},
  author={Sakour, Ibraheem and Hu, Huosheng},
  booktitle={2016 8th Computer Science and Electronic Engineering (CEEC)},
  pages={112--117},
  year={2016},
  organization={IEEE}
}

@article{DBLP:journals/concurrency/FischerGJRWR21,
  author       = {Joel E. Fischer and
                  Chris Greenhalgh and
                  Wenchao Jiang and
                  Sarvapali D. Ramchurn and
                  Feng Wu and
                  Tom Rodden},
  title        = {In-the-loop or on-the-loop? Interactional arrangements to support
                  team coordination with a planning agent},
  journal      = {Concurr. Comput. Pract. Exp.},
  volume       = {33},
  number       = {8},
  year         = {2021},
  url          = {https://doi.org/10.1002/cpe.4082},
  doi          = {10.1002/cpe.4082},
  bibsource    = {dblp computer science bibliography, https://dblp.org}
}

@inproceedings{DBLP:conf/icse/Lemos20,
  author       = {Rog{\'{e}}rio de Lemos},
  _editor       = {Shinichi Honiden and
                  Elisabetta Di Nitto and
                  Radu Calinescu},
  title        = {Human in the loop: what is the point of no return?},
  booktitle    = {{SEAMS} '20: {IEEE/ACM} 15th International Symposium on Software Engineering
                  for Adaptive and Self-Managing Systems, Seoul, Republic of Korea,
                  29 June - 3 July, 2020},
  pages        = {165--166},
  publisher    = {{ACM}},
  year         = {2020},
  url          = {https://doi.org/10.1145/3387939.3391597},
  doi          = {10.1145/3387939.3391597},
  timestamp    = {Thu, 24 Sep 2020 12:13:39 +0200},
  biburl       = {https://dblp.org/rec/conf/icse/Lemos20.bib},
  bibsource    = {dblp computer science bibliography, https://dblp.org}
}

@inproceedings{DBLP:conf/seams/LiCGSJ21,
  author       = {Nianyu Li and
                  Javier C{\'{a}}mara and
                  David Garlan and
                  Bradley R. Schmerl and
                  Zhi Jin},
  title        = {Hey! Preparing Humans to do Tasks in Self-adaptive Systems},
  booktitle    = {16th International Symposium on Software Engineering for Adaptive
                  and Self-Managing Systems, SEAMS@ICSE 2021, Madrid, Spain, May 18-24,
                  2021},
  pages        = {48--58},
  publisher    = {{IEEE}},
  year         = {2021},
  url          = {https://doi.org/10.1109/SEAMS51251.2021.00017},
  doi          = {10.1109/SEAMS51251.2021.00017},
  bibsource    = {dblp computer science bibliography, https://dblp.org}
}

@inproceedings{DBLP:conf/icse/CalikliLBRDPSLN16,
  author    = {G{\"{u}}l {\c{C}}alikli and
               Mark Law and
               Arosha K. Bandara and
               Alessandra Russo and
               Luke Dickens and
               Blaine A. Price and
               Avelie Stuart and
               Mark Levine and
               Bashar Nuseibeh},
  title     = {Privacy dynamics: learning privacy norms for social software},
  booktitle = {Proceedings of the 11th International Symposium on Software Engineering
               for Adaptive and Self-Managing Systems, SEAMS@ICSE 2016, Austin, Texas,
               USA, May 14-22, 2016},
  pages     = {47--56},
  publisher = {{ACM}},
  year      = {2016},
  url       = {https://doi.org/10.1145/2897053.2897063},
  doi       = {10.1145/2897053.2897063},
  bibsource = {dblp computer science bibliography, https://dblp.org}
}

@article{heimlich2008understanding,
  title={Understanding behavior to understand behavior change: A literature review},
  author={Heimlich, Joe E and Ardoin, Nicole M},
  journal={Environmental education research},
  volume={14},
  number={3},
  pages={215--237},
  year={2008},
  publisher={Taylor \& Francis}
}

@article{sebanz2006joint,
  title={Joint action: bodies and minds moving together},
  author={Sebanz, Natalie and Bekkering, Harold and Knoblich, G{\"u}nther},
  journal={Trends in cognitive sciences},
  volume={10},
  number={2},
  pages={70--76},
  year={2006},
  publisher={Elsevier}
}

@inproceedings{camara2015reasoning,
  title={Reasoning about human participation in self-adaptive systems},
  author={C{\'a}mara, Javier and Moreno, Gabriel and Garlan, David},
  booktitle={2015 IEEE/ACM 10th International Symposium on Software Engineering for Adaptive and Self-Managing Systems},
  pages={146--156},
  year={2015},
  organization={IEEE}
}

@inproceedings{baresi2024conceptual,
  title={A Conceptual Framework for Quality Assurance of LLM-based Socio-critical Systems},
  author={Baresi, Luciano and Camilli, Matteo and Dolci, Tommaso and Quattrocchi, Giovanni},
  booktitle={Proceedings of the 39th IEEE/ACM International Conference on Automated Software Engineering},
  pages={2314--2318},
  year={2024}
}

@inproceedings{DBLP:conf/icse/RaufLLTNR20,
  author    = {Irum Rauf and
               Dirk van der Linden and
               Mark Levine and
               John N. Towse and
               Bashar Nuseibeh and
               Awais Rashid},
  title     = {Security but not for security's sake: The impact of social considerations
               on app developers' choices},
  booktitle = {{ICSE} '20: 42nd International Conference on Software Engineering,
               Workshops, Seoul, Republic of Korea, 27 June - 19 July, 2020},
  pages     = {141--144},
  publisher = {{ACM}},
  year      = {2020},
  url       = {https://doi.org/10.1145/3387940.3392230},
  doi       = {10.1145/3387940.3392230},
  bibsource = {dblp computer science bibliography, https://dblp.org}
}

@article{cacm23,
  author       = {Dhaminda B. Abeywickrama and
                  Amel Bennaceur and
                  Greg Chance and
                  Yiannis Demiris and
                  Anastasia Kordoni and
                  Mark Levine and
                  Luke Moffat and
                  Luc Moreau and
                  Mohammad Reza Mousavi and
                  Bashar Nuseibeh and
                  Subramanian Ramamoorthy and
                  Jan Oliver Ringert and
                  James Wilson and
                  Shane Windsor and
                  Kerstin Eder},
  title        = {On Specifying for Trustworthiness},
  journal      = {Commun. {ACM}},
  volume       = {67},
  number       = {1},
  pages        = {98--109},
  year         = {2024},
  url          = {https://doi.org/10.1145/3624699},
  doi          = {10.1145/3624699},
  bibsource    = {dblp computer science bibliography, https://dblp.org}
}

@article{bertinoro2023,
author = {Weyns, Danny and Calinescu, Radu and Mirandola, Raffaela and Tei, Kenji and Acosta, Maribel and Bennaceur, Amel and Boltz, Nicolas and Bures, Tomas and Camara, Javier and Diaconescu, Ada and Engels, Gregor and Gerasimou, Simos and Gerostathopoulos, Ilias and Getir Yaman, Sinem and Grassi, Vincenzo and Hahner, Sebastian and Letier, Emmanuel and Litoiu, Marin and Marsso, Lina and Musil, Angelika and Musil, Juergen and Nunes Rodrigues, Genaina and Perez-Palacin, Diego and Quin, Federico and Scandurra, Patrizia and Vallecillo, Antonio and Zisman, Andrea},
title = {Towards a Research Agenda for Understanding and ManagingUncertainty in Self-Adaptive Systems},
year = {2023},
issue_date = {October 2023},
publisher = {Association for Computing Machinery},
address = {New York, NY, USA},
volume = {48},
number = {4},
issn = {0163-5948},
url = {https://doi-org.libezproxy.open.ac.uk/10.1145/3617946.3617951},
doi = {10.1145/3617946.3617951},
journal = {SIGSOFT Softw. Eng. Notes},
month = {oct},
pages = {20–36},
numpages = {17}
}

@inproceedings{AlmalikiFBPA14,
  author       = {Malik Almaliki and
                  Funmilade Faniyi and
                  Rami Bahsoon and
                  Keith Phalp and
                  Raian Ali},
  _editor       = {Camille Salinesi and
                  Inge van de Weerd},
  title        = {Requirements-Driven Social Adaptation: Expert Survey},
  booktitle    = {Requirements Engineering: Foundation for Software Quality - 20th International
                  Working Conference, {REFSQ} 2014, Essen, Germany, April 7-10, 2014.
                  Proceedings},
  series       = {Lecture Notes in Computer Science},
  volume       = {8396},
  pages        = {72--87},
  publisher    = {Springer},
  year         = {2014},
  url          = {https://doi.org/10.1007/978-3-319-05843-6\_6},
  doi          = {10.1007/978-3-319-05843-6\_6},
  timestamp    = {Tue, 14 May 2019 10:00:39 +0200},
  biburl       = {https://dblp.org/rec/conf/refsq/AlmalikiFBPA14.bib},
  bibsource    = {dblp computer science bibliography, https://dblp.org}
}

@article{DBLP:journals/cacm/Kress-GazitEHAA21,
  author       = {Hadas Kress{-}Gazit and
                  Kerstin Eder and
                  Guy Hoffman and
                  Henny Admoni and
                  Brenna Argall and
                  R{\"{u}}diger Ehlers and
                  Christoffer Heckman and
                  Nils Jansen and
                  Ross A. Knepper and
                  Jan Kret{\'{\i}}nsk{\'{y}} and
                  Shelly Levy{-}Tzedek and
                  Jamy Li and
                  Todd D. Murphey and
                  Laurel D. Riek and
                  Dorsa Sadigh},
  title        = {Formalizing and guaranteeing human-robot interaction},
  journal      = {Commun. {ACM}},
  volume       = {64},
  number       = {9},
  pages        = {78--84},
  year         = {2021},
  url          = {https://doi.org/10.1145/3433637},
  doi          = {10.1145/3433637},
  bibsource    = {dblp computer science bibliography, https://dblp.org}
}

@article{DBLP:journals/corr/abs-1903-04771,
  author       = {Danny Weyns and
                  Nelly Bencomo and
                  Radu Calinescu and
                  Javier C{\'{a}}mara and
                  Carlo Ghezzi and
                  Vincenzo Grassi and
                  Lars Grunske and
                  Paola Inverardi and
                  Jean{-}Marc J{\'{e}}z{\'{e}}quel and
                  Sam Malek and
                  Raffaela Mirandola and
                  Marco Mori and
                  Giordano Tamburrelli},
  title        = {Perpetual Assurances for Self-Adaptive Systems},
  journal      = {CoRR},
  volume       = {abs/1903.04771},
  year         = {2019},
  url          = {http://arxiv.org/abs/1903.04771},
  eprinttype    = {arXiv},
  eprint       = {1903.04771},
  bibsource    = {dblp computer science bibliography, https://dblp.org}
}

@article{DBLP:journals/corr/abs-2307-03697,
  author       = {Sinem Getir Yaman and
                  Ana Cavalcanti and
                  Radu Calinescu and
                  Colin Paterson and
                  Pedro Ribeiro and
                  Beverley Townsend},
  title        = {Specification, Validation and Verification of Social, Legal, Ethical,
                  Empathetic and Cultural Requirements for Autonomous Agents},
  journal      = {CoRR},
  volume       = {abs/2307.03697},
  year         = {2023},
  url          = {https://doi.org/10.48550/arXiv.2307.03697},
  doi          = {10.48550/ARXIV.2307.03697},
  eprinttype    = {arXiv},
  eprint       = {2307.03697},
  bibsource    = {dblp computer science bibliography, https://dblp.org}
}

@article{cleland23,
author = {Cleland-Huang, Jane and Chambers, Theodore and Zudaire, Sebastian and Chowdhury, Muhammed Tawfiq and Agrawal, Ankit and Vierhauser, Michael},
title = {Human-Machine Teaming with Small Unmanned Aerial Systems in a MAPE-K Environment},
year = {2023},
publisher = {Association for Computing Machinery},
address = {New York, NY, USA},
issn = {1556-4665},
url = {https://doi-org.libezproxy.open.ac.uk/10.1145/3618001},
doi = {10.1145/3618001},
note = {Just Accepted},
journal = {ACM Trans. Auton. Adapt. Syst.},
month = {sep}
}

@inproceedings{Lo23,
  author       = {David Lo},
  title        = {Trustworthy and Synergistic Artificial Intelligence for Software Engineering:
                  Vision and Roadmaps},
  booktitle    = {ICSE-FoSE},
  pages        = {69--85},
  publisher    = {{IEEE}},
  year         = {2023}
}

@article{LoLLM24,
  author       = {Junda He and
                  Christoph Treude and
                  David Lo},
  title        = {LLM-Based Multi-Agent Systems for Software Engineering: Vision and
                  the Road Ahead},
  journal      = {CoRR},
  volume       = {abs/2404.04834},
  year         = {2024}
}

@inproceedings{HassanLRGC00TOL24,
  author       = {Ahmed E. Hassan and
                  Dayi Lin and
                  Gopi Krishnan Rajbahadur and
                  Keheliya Gallaba and
                  Filipe Roseiro C{\^{o}}go and
                  Boyuan Chen and
                  Haoxiang Zhang and
                  Kishanthan Thangarajah and
                  Gustavo Ansaldi Oliva and
                  Jiahuei (Justina) Lin and
                  Wali Mohammad Abdullah and
                  Zhen Ming (Jack) Jiang},
  title        = {Rethinking Software Engineering in the Era of Foundation Models: {A}
                  Curated Catalogue of Challenges in the Development of Trustworthy
                  FMware},
  booktitle    = {{SIGSOFT} {FSE} Companion},
  pages        = {294--305},
  publisher    = {{ACM}},
  year         = {2024}
}

@INPROCEEDINGS{grigore:2013,
	author={Grigore, Elena Corina and Eder, Kerstin and Pipe, Anthony G. and Melhuish, Chris and Leonards, Ute},
	booktitle={Proc. of the 2013 IEEE/RSJ International Conference on Intelligent Robots and Systems}, 
	title={Joint action understanding improves robot-to-human object handover}, 
	year={2013},
	volume={},
	number={},
	pages={4622-4629},
	doi={10.1109/IROS.2013.6697021}
}

@inproceedings{DBLP:conf/iscram/JohnsonCSMP21,
  author       = {Kenneth Johnson and
                  Javier C{\'{a}}mara and
                  Roopak Sinha and
                  Samaneh Madanian and
                  Dave Parry},
  _editor       = {Anouck Adrot and
                  Rob Grace and
                  Kathleen A. Moore and
                  Christopher W. Zobel},
  title        = {Towards Self-Adaptive Disaster Management Systems},
  booktitle    = {18th International Conference on Information Systems for Crisis Response
                  and Management, {ISCRAM} 2021, Blacksburg, VA, USA, May 2021},
  pages        = {49--61},
  publisher    = {{ISCRAM} Digital Library},
  year         = {2021},
  url          = {https://idl.iscram.org/show.php?record=2312},
  timestamp    = {Thu, 10 Nov 2022 16:58:41 +0100},
  biburl       = {https://dblp.org/rec/conf/iscram/JohnsonCSMP21.bib},
  bibsource    = {dblp computer science bibliography, https://dblp.org}
}

@inproceedings{DBLP:conf/icse/LiAKG20,
  author       = {Nianyu Li and
                  Sridhar Adepu and
                  Eunsuk Kang and
                  David Garlan},
  _editor       = {Shinichi Honiden and
                  Elisabetta Di Nitto and
                  Radu Calinescu},
  title        = {Explanations for human-on-the-loop: a probabilistic model checking
                  approach},
  booktitle    = {{SEAMS} '20: {IEEE/ACM} 15th International Symposium on Software Engineering
                  for Adaptive and Self-Managing Systems, Seoul, Republic of Korea,
                  29 June - 3 July, 2020},
  pages        = {181--187},
  publisher    = {{ACM}},
  year         = {2020},
  url          = {https://doi.org/10.1145/3387939.3391592},
  doi          = {10.1145/3387939.3391592},
  timestamp    = {Thu, 24 Sep 2020 12:13:39 +0200},
  biburl       = {https://dblp.org/rec/conf/icse/LiAKG20.bib},
  bibsource    = {dblp computer science bibliography, https://dblp.org}
}

@inproceedings{DBLP:conf/seams/PurandareSICC23,
  author       = {Salil Purandare and
                  Urjoshi Sinha and
                  Md Nafee Al Islam and
                  Jane Cleland{-}Huang and
                  Myra B. Cohen},
  title        = {Self-Adaptive Mechanisms for Misconfigurations in Small Uncrewed Aerial
                  Systems},
  booktitle    = {18th {IEEE/ACM} Symposium on Software Engineering for Adaptive and
                  Self-Managing Systems, {SEAMS} 2023, Melbourne, Australia, May 15-16,
                  2023},
  pages        = {169--180},
  publisher    = {{IEEE}},
  year         = {2023},
  url          = {https://doi.org/10.1109/SEAMS59076.2023.00030},
  doi          = {10.1109/SEAMS59076.2023.00030},
  bibsource    = {dblp computer science bibliography, https://dblp.org}
}
\balance

\end{document}